# Congestion Control Approach based on Effective Random Early Detection and Fuzzy Logic

**Maimuna Khatari and Ghassan Samara**

Internet Technology Department, Faculty of Information Technology, Zarqa University, Zarqa, Jordan

Mkhatari, Gsamara@zu.edu.jo

**Abstract: Congestion** in router buffer increases the **delay** and **packet loss**. Active Queue Management (**AQM**) methods are able to detect congestion in early stage and control it by packet dropping. Effective Random Early Detection (ERED) method, among many other AQM methods, gives a good performance in detect and control congestion and preserve **packet loss**. However, the ERED neglect the **delay** factor, which is effect the performance of the network. Moreover, ERED has a real parameterization problem. Several parameters have to be initialized to optimal values to obtain satisfactory results. This paper proposed an extended ERED method that considers the delay in its process and combines the extended ERED method with a Fuzzy Inference Process that eases the problem of parameter initialization. The results show that the parametric-based form of the proposed work gives a better performance results, according to the performance measures, delay, dropping and packet loss. The loss has been enhanced by 10-100%. Delay has been enhanced by 30-60%. The performance of the fuzzy-based form of the proposed method is better than the parametric-based form and ERED in terms of delay and packet loss.

**1. Introduction**

Data communicated and transmitted, from *host-to-host* via the internet, is passed through various network topologies and intermediated by *links* and *routers*, which form the main network resources. **Congestion** referred to the increment of the data rate in a way that the network resources cannot handle it. A host sending data over TCP protocol avoids congestion by a widowing mechanism that adjusts the sending rates based on the status of the network, estimated by the time and amount of the received acknowledgements. Other protocols, such as UDP and protocols that are used in wireless and mobile ad-hock networks, do not adhere to such mechanism. Transmitting using such protocols and even with the TCP protocol may lead to **congestion** at the network *routers* [1-5].

Active Queue Management (*AQM*) is a well known set of methods for route-based congestion control. **AQM** methods detect and control congestion in early stage and start dropping packets early aiming at reducing *packet loss* and *delay*. These methods depend on calculating dropping probability, *Dp*, for each arrival packet, in order to prevent congestion. *Dp* is then used to make a decision of dropping or accommodating the packet. Examples of the existing **AQM** methods are: Random Early Detection (RED), Adaptive RED (ARED), Gentle RED (GRED), Adaptive GRED (AGRED), Dynamic GRED (DGRED), BLUE, dynamic threshold based BLUE (DT BLUE), GREEN, Adaptive CHOKE, ERED, Subsidized RED (SubRED) and others [6].

RED [7], the foundation of all AQM methods, and many other AQM methods, uses *average queue length (avg)* as indicator of the queue status. The *avg* is evaluated in multiple levels that are bounded by multiple thresholds. *minThreshold* is a position in the buffer, if not exceeded by *avg*, the buffer can be considered to be of fair amount of packets in the queue. While, *maxThreshold* parameter is another position, if reached or exceeded by *avg*, may be an indication of congestion. Accordingly, *Dp* increases when the queued packets reach *maxThreshold*.

RED avoids congestion by marking or dropping packets at early stage with a probability value *Dp*. The dropping probability is a function of *avg*. Subsequently, when the *avg* is less than *minThreshold*, RED drops no packets. While, all packets are dropped if *avg* exceeds *maxThreshold*. The actual congestion control is implemented when *avg* is between *minThreshold* and *maxThreshold*. The advantages of RED are as following:

- RED predicts congestion in early stage as it responses, by dropping packets, to the increment of packet queuing in the buffer.
- RED avoids global synchronization by dropping packets randomly.
- RED avoids dropping packets unnecessarily when a short heavy traffic is presented (false congestion) as it deals with *avg* as congestion indicator not *q*. *avg* dose not increased rapidly when false congestion occurs.

RED has been successfully adopted by the Internet Engineering Task Force (IETF) in RFC 2309. However, as the technology has been evolving rapidly, the problems of RED have been noticed, which can be summarized as following [8]:

- RED suffers from insensitivity to current queue status, in sudden congestion, RED slowly adapts and results in *packet loss*, as it uses *avg* instead of *q*. Notice that considering *q* will lead to drop packets unnecessarily when a short heavy traffic presented.
- The parameter initialization problem of the RED which affects the performance of RED.
- RED may harm the *delay* as it does not take into consideration the arrival and departure rates (traffic-load amount).

The Effective RED (ERED) [5] was proposed to overcome the insensitivity to current queue size [5]. ERED uses the instantaneous queue size (*q*) as congestion indicator side by side with the *avg*. Congestion is controlled based on the following scenarios: When *avg* is between *minThreshold* and *maxThreshold* and *q* is greater than *minThreshold*, ERED drops arrival packets with *Dp* as calculated in RED. While, when *avg* is less than *minThreshold* and *q* is less than 1.75* *maxThreshold*, ERED drops arrival packets with a high *Dp* value. In the rest of the scenarios ERED acts like RED does. Subsequently, ERED solves the problem of insensitivity to current queue status, in sudden congestion. However, the ERED's drawbacks are the parameterization. Moreover, ERED does not takes into consideration the *delay* in the calculation of *Dp*.

This paper proposed an extended ERED method that considers the *delay* in its process and combines the extended ERED method with a Fuzzy Inference Process (FIP) that eases the problems of parameter initialization and parameter dependency.

**2. Related Works**

To address the delay issue, Adaptive Virtual Queue (AVQ) AVQ [9] maintains a virtual queue whose capacity is less than the actual capacity of the real queue. The number of packet queued in the virtual queue is synchronized with the number in the real one. Packets are dropped from the real queue when the virtual buffer is overflowed. A virtual capacity of the connected links are also maintained. The utilization of real links (arrival and departure) are maintained based on the virtual capacity using a desired utilization parameter γ. AVQ is responsive to changes in network load and is able to maintain a small queue length even when network load is increased [9]. Considering the load in the AQM enhances the delay. However, there is tradeoff between queuing delay and queue utilization. Intuitively, a larger buffer leads to higher utilizations of the resources, but it also results in more queuing delays.

Subsequently, hybrid approaches, such as REM [10] and PI [11] that balance between the two factors were emerged. Random Exponential Marking (REM) [10], uses two congestion indicators that are the transmission rate (*R*) and queue size (*q*). These indicators are combined to come out with the total price property. The price property is used to find out the *Dp*. By using *R*, REM aims at stabilizing the transmission rates of the sources at the link capacity (L) and by using *q*, REM aims at stabilizing the queue size at a certain level (target queue length (Tql)). Subsequently, the price property which controls *Dp,* depends on the rate mismatch and the queue mismatch. The rate mismatch corresponds to the differences between the transmission rates and the link capacity, and the queue mismatch equals the differences between the queue length and the target queue length [10, 12]. Unfortunately, REM has limitations such as, parameterization and low throughput when the traffic is high. Similarly, Proportional Integral Controller (PI) [11] uses, as congestion indicators, *q* and loss probability which is estimated based on the load. While these methods address delay, they depend mainly on the current queue status, *q*. *q*, as congestion indicator, leads to drop packets unnecessarily in false congestion status as discussed earlier. Moreover, these methods required massive parameter initialization.

To deal with the parameterization problem, several fuzzy-based AQM methods were proposed. With AQM methods, FL is used to calculate *Dp*. Fuzzy Explicit Marking (FEM) [13-15] is AQM method that replaces the utilized thresholds in RED and other AQM methods with a fuzzy inference process (FIP). FEM calculates *Dp* as an output linguistic variable based on two input linguistic variables, that are, the changing rate in the buffer (CRB) and *q* as parallel to using *avg* in RED [14, 16]. REDD1 [17] is a fuzzy-based method that was built on top of RED method as well. REDD1 relies on two input linguistic variables, which are: the average queue length (*avg*) and packet loss (*$P_L$*), to generate one output linguistic variable (*Dp*). Fuzzy BLUE (FB) extends BLUE method using fuzzy logic theory [14, 18-20]. FB depends on two input linguistic variables, buffer occupancy (BO) and packet loss (PL) for calculating one output, which is *Dp* . As noted, these fuzzy-based methods were proposed as extensions to the existing AQM methods. Subsequently, while these fuzzy-based methods solve the parameter initialization problem, they inherit the limitation of the original methods as discussed earlier. Overall, there is no single method addresses all the problems inherited from RED, the basic of the all AQM methods.

## 3. Proposed Work

The proposed hybrid-ERED uses three congestion indicators, which are: *q*, *avg* and *$D_{Esti}$*. *$D_{Esti}$* will be calculated based on the *arrival rate* and *departure rate*. *Dp* will be calculated to be increased/decreased linearly with the increment/decrement of the value of the congestion indicators. In the hybrid-ERED, a few scenarios are considered that are drawn based on the status of the indicators according to some thresholds.

### 3.1 Delay Calculation

The overall network delay is composed of different types of delays, these are: processing delay, transmission delay, propagation delay and queuing delay. The queuing delay can be calculated according to various laws. In this paper, a derivation of the Little's law [21], which has a fair estimation of the time that packet spends in the queue is adopted. As such, estimated delay *$D_{Esti}$*, is calculated as given in Equation 1.

$$D_{Esti} = (1 - Dep_{Esti}) + Arr_{Esti} \qquad (1)$$

The estimated delay, *$D_{Esti}$* is proportional to the increment of the arrival rate and inverse proportional to the departure rate. The estimated arrival rate is low-pass filter of the average arrival rates over the previous and the current packets arrival rates. The arrival rate is calculated as given in Equation 2.

$$Arr_{Esti} = Arr_{p-1}(1 - w_{arr}) + Arr_p(w_{arr}) \qquad (2)$$

where p is the time value, *$Arr_{p-1}$* is the value calculated previously for the arrival rate, *$Arr_p$* is the number of packets arrived in current time and *$w_{arr}$* is weight parameter. The estimated departure rate is calculated similarly as given in Equation 3.

$$Dep_{Esti} = Dep_{p-1}(1 - w_{dep}) + Dep_p(w_{dep}) \qquad (3)$$

The utilized low pass filter, that is used for calculating the estimated arrival and departure, gives more contribution to the previous rates. This is implemented by selecting values for the weight parameters *$w_{arr}$* and *$w_{dep}$* below 0.5. Subsequently, the calculated output for the estimated arrival and departure, is changed gradually when the rate changes overtime.

The value of *$D_{Esti}$* is optimal, according to Equation 1 and with reference to Equation 2 and Equation 3, when the rate of arrival is equal to the rate of departure and only a single packet is queued in the buffer. The value of one is the edge between positive/high delay and negative/no-delay. Positive/high delay has a value above one, which is occur when the arrival rate is more than the departure rate and the queue is not empty. Negative/no-delay delay has a value below one, which is occur when the departure rate is higher than the arrival rate and the queue size is low. ***D_{Esti}*** is proportional with the term (***Arr_{Esti}/Dep_{Esti}***) as given in Fig. 1.

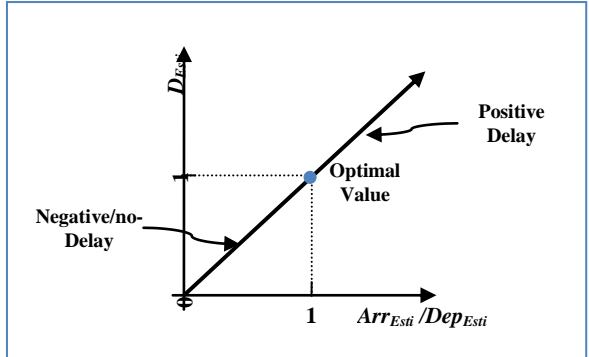

**Fig. 1.** $D_{Esti}$ vs. ($Arr_{Esti}/Dep_{Esti}$)

### 3.2 Hyprid-ERED Mechanism

In the proposed Hyprid-ERED, ***avg*** and ***q*** indicators are tight up with thresholds. These indicators according to their values with respect to the thresholds estimate the status of the congestion. Based on the values of ***avg*** and ***q,*** Hyprid-ERED either drops no packets, drops packets based on ***Dp*** that is calculated using Equation 4, drops packets based on ***Dp*** that is calculated using Equation 5, which generates values below the values generated in Equation 4, subsequently drop less packets or drops all arrival packets.

$$D_p' = \frac{max_p(aql - min_{th})}{max_{th} - min_{th}}$$

$$, D_p = \frac{D_p'}{1 - count * D_p'} + w_D(D_{Esti}) \quad (4)$$

$$D_p = \frac{D_{max}}{1 - count * D_{max}} \quad (5)$$

Equation 5 is borrowed from ERED as it is. This Equation drops less packets than Equation 4 and it is used in non-severe congestion. Thus, adjusting this Equation is not valuable. The first part of Equation 4 is also borrowed from ERED and is coupled with the second part that has been developed in the proposed Hyprid-RED to consider the delay. The parameter $w_D$ is a weight value determines the contribution of the delay status in the network to the overall calculated ***Dp*** value.

### 3.3 Algorithm

The Hybrid-ERED algorithm is given below in Algorithm 1.

**Algorithm 1: Hybrid-ERED**

1  INITIALIZATION:
2     ***avg***:= 0
3     ***count***:= 1
4  FOR EACH arrival packet
5     CALCULATE new ***avg*** as follows:
6        IF   ***q***==0 THEN ***avg***:=(1-w)$^{f(time- q\_time)}$ *
7  ***avg***
8        IF ***q***!= 0 THEN ***avg***:= (1-w)* ***avg*** + w ****q***
9     CALCULATE ***D*** and its related parameters
10 as follows:
11       IF   ***q***:= FULL THEN arr$_p$ = arr$_{p-1}$ and
12 dep$_p$:= (1-w$_{dep}$)* dep$_{p-1}$ + w$_{dep}$ * #departed$_p$
13       IF ***q***!= FULL THEN
14          arr$_p$:= (1-w$_{arr}$)* arr$_{p-1}$ + w$_{arr}$ * #arrived$_p$
15 and   dep$_p$:= (1-w$_{dep}$)* dep$_{p-1}$ + w$_{dep}$ *
16 #departed$_p$
17       ***D***:= (1/ dep$_p$) * arr$_p$ *q
18    CALCULATE    ***Dp***    and    its    related
19 parameters,    and    implements    packet
20 dropping, as:
21    if (min$_{th2}$ ≤ ***avg*** < max $_{th3}$) && (***q***≥min $_{th2}$)
22       increase count
23       ***Dp'***= max$_p$* (***avg***-min $_{th}$)/(max $_{th}$-min
24 $_{th}$)
25       ***Dp*** = ***Dp'***/ (1-count* ***Dp'***) + w$_d$(D)
26       with probability ***Dp***
27          drop packet
28          ***count*** := 0
29    else if (***avg*** < min $_{th2}$) && (***q***>max $_{th2}$)
30       Calculate ***Dp*** = max$_p$/(1-count* max$_p$)
31       with probability ***Dp***
32          drop packet
33          count := 0
34    else if (***avg*** ≥ max $_{th3}$)
35       Drop packet
36       Count = 0
37    else
38       Count = -1
39    When q==0
40       q_time=time

Lines   1-3   provide   the   necessary initialization, in which the parameter ***avg*** is set to 0 and ***count*** to -1. Lines 4-32 are repeated with each arrival packet. Mainly, four major process is implemented with each arrival packet, there are:
- Calculate the value of ***avg***, in lines 5-7 as follows: Case 1 in line 6, if ***q*** is 0, ***q*** is the instance size of the buffer, then ***avg*** is calculated based on the function powered by the idle time. The more idle time (the longer the buffer remains empty), the larger the value of the function and the lower the value of *avg*, as given in Equation 6. Case 2 in line 7, if *q* is not 0, *avg* is calculated as low-pass filter of the average queue size, as given in Equation 7.

$$\text{avg}_p = aql_{p-1}(1 - w_{aql})^{f(idelTime)} \quad (6)$$
$$\text{avg}_p = aql_{p-1}(1 - w_{aql}) + q_p(w_{aql}) \quad (7)$$

- Calculate the value of $D_{Esti}$, lines 8-12, as follows: Case 1 in line 9, if the *q* is full, then packets will be dropped and no packets will be arrived, subsequently, the estimated arrival $Arr_{Esti}$, cannot be updated and it remains as calculated before the buffer overflowed, while the estimated departure $Dep_{Esti}$, is calculated as given in Equation 3 and discussed earlier. Case 2 in lines 10-11, if *q* is not full, both the estimated arrival $Arr_{Esti}$, and estimated departure $Dep_{Esti}$, are calculated as given in Equation 2, Equation 3, which are discussed earlier.
- Calculate the value of *Dp* and implements packet dropping or queuing, lines 13-31, as follows: Case 1 in lines 14-20, if *avg* between $min_{th2}$ and $max_{th3}$, and *q* above $min_{th2}$, the dropping probability, *Dp,* is calculated as given in Equation 4. If the packet is dropped, the value of the variable count is set to 0, as given in line 20. Case 2 in lines 21-25, if *avg* is less than $min_{th2}$ and *q* above $max_{th2}$, the dropping probability, *Dp*, is calculated as given in Equation 5. If the packet is dropped, the value of the variable count is set to 0, the variable count is set to 0. Case 3 in lines 26-28, if *avg* is more than $max_{th3}$, the dropping probability *Dp* is set to 1, the packet subject matter is dropped and the value of the variable count is set to 0. Case 4 in lines 29-30, if *avg* is between $min_{th2}$ and $max_{th3}$ and *q* vale is less than $min_{th2}$ or if *avg* is less than $min_{th2}$ and *q* vale is between $min_{th2}$ and $max_{th2}$, the dropping probability *Dp* is set to 0 and the value of the variable count is set to -1.
- Storing the starting of idle time, lines 32-33, the idle time starts when the buffer gets empty.

Generally, the proposed method control congestion by calculating a *Dp* value based on some input measures that are estimated from the current buffer and traffic statues, there parameters are: Queue size *q*, Average Queue Length *avg*, Estimated Arrival Rate $Arr_{Esti}$, Estimated Departure Rate $Dep_{Esti}$, Estimated Delay $D_{Esti}$, A fixed value $max_p$ and *count* since the last dropped packet (-1 or 0) *count*. Mainly, dropping is controlled when *q*, *avg* values are between some thresholds that are: $min_{th}$, $min_{th2}$, $max_{th}$, $max_{th2}$, $max_{th3}$. The utilized, measures, parameters and thresholds are identified as follows:

1. Queue (*q*): The instance number of packets that are queued in the router buffer.
2. Average Queue Length (*avg*): The average number of packets over a time window, which is calculated using low-pass filter.
3. Estimated Arrival Rate ($Arr_{Esti}$): The average number of packets received in a time window, which is calculated as mean value.
4. Estimated Departure Rate ($Dep_{Esti}$): The average number of packets departure in a time window, which is calculated as mean value.
5. Estimated Delay ($D_{Esti}$): reflects the amount of time units the packets stayed in the buffer.
6. The count since the last dropped packet (*count*): either packet has been dropped just now (0) or not (-1).

The thresholds, are:
1. $min_{th}$: A fixed value that refers to a position in the router buffer, if not exceeded by the queued packets, according to *avg* and *q*, no dropping is necessary as the queue will be considered of fair size.
2. $min_{th2}$: A value that refers to a position in the router buffer of unite situation. $min_{th2}$: is not a fixed value and it is calculated as it is calculated as given in Equation 8.
3. $max_{th}$: A fixed value that refers to a position in the router buffer, if exceeded by the queued packets, according to *avg* and *q*, the arrived packets should be dropped as the queue will be considered to be in a risk.
4. $max_{th2}$: A value that refers to a position in the router buffer of sever situation. $max_{th2}$ is not a fixed value and it is calculated as it is calculated as given in Equation 9.
5. $max_{th3}$: A value that refers to a position in the router buffer, if exceeded by the queued packets, according to *avg* and *q*, the arrived packets should be dropped. This stage is more risky than previous one. $max_{th3}$ is not a fixed value and it is calculated as it is calculated as given in Equation 10. These parameters and equations have been borrowed from ERED.

$$min_{th2} = \frac{max_{th} + min_{th}}{2} + min_{th} \quad (8)$$
$$max_{th2} = 1.75 * max_{th} \quad (9)$$
$$max_{th3} = 2 * max_{th} \quad (10)$$

*Dp* is calculated based on the status of the buffer and with reference to the parameters identified previously based on the following scenarios:
1. When *avg* is between $min_{th2}$ and $max_{th3}$ and *q* vale is less than $min_{th2}$ or if the *avg* is less than $min_{th2}$ and *q* vale is between $min_{th2}$ and $max_{th2}$, no packets are dropped (e.g.: *Dp*=0.0).
2. When *avg* above $max_{th3}$, all arrived packets are dropped (e.g.: *Dp*=1.0).
3. When *avg* between $min_{th2}$ and $max_{th3}$ and *q* above $min_{th2}$, packets are dropped randomly with probability *Dp*.
4. When *avg* is less than $min_{th2}$ and *q* above $max_{th2}$, packets are dropped randomly with probability *Dp*. However, with probability value less than the one calculated in the previous case.

### 3.4 Fuzzy-based

The proposed hyprid-ERED is wrapped with fuzzy inference process to ease the parameter initialization problem, in this section. Specifically, the thresholds and the $max_p$ parameters' values should be eliminated. The proposed method relies on two input linguist variables that are *avg* and *q*. The output is an initial *Dp* value. As a FIP-based, the proposed method, will be implemented in sequential process that are, the fuzzification, rules evaluation, aggregation and defuzzification. The process of implementing Fuzzy-based hybrid ERED is given in Fig. 2.

### 3.4.1 Fuzzification

In the *Fuzzification* step, the input crisp for the linguist variables *avg* and *q* values are converted into equivalent linguistic terms in some fuzzy sets. The fuzzy sets are determined as: *q*: {empty, low, moderate, full}, *avg*:{low, moderate, high}, *Dp*: {zero, low, medium, high}. The membership function can be formulated for the input and output linguistic variables, using well-established approaches of trapezoidal [20] or triangular [14].

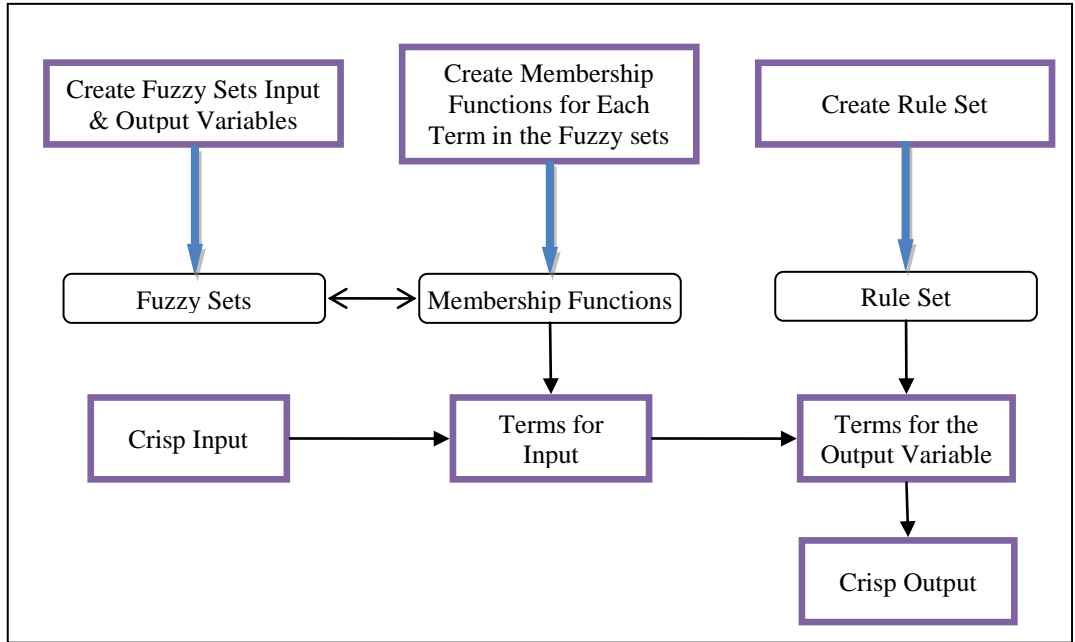

**Fig. 2.** Flowchart of Fuzzy Hybrid ERED

For the simplicity, the triangular membership function was selected. The membership function should be trained to determines the boundary and the slop of each linguistic term. As the training process implemented, each crisp value can be converted into linguistic term in its corresponding fuzzy set. The input crisp value can be converted into multiple linguistic term, each with a probability value.

### 3.4.2 Rules Evaluation

In the *Rules Evaluation* step, the rules utilized to obtain the linguistic terms for the output variable from the linguistic terms of the input variables, are determined. Examples of the fuzzy rules are given in Table 1 [13]. More rules are added to be tested as given in Table 2. Table 2, gives the rules in tables, where rows and columns header represent input terms, while the cells filled in with the output term, for simplicity. However, all these rules will be tested and verified, the final set of rules will be obtained when the experiments are implemented.

### 3.4.3 Aggregation

The next step is to *aggregate the output of all rules*. If the inputs can be anticipated on multiple rules, then multiple outputs will be obtained. Each of these will also have a crisp value which is obtained in this step. This step is start with obtain a crisp value for each output in which the input can be anticipated. The crisp value of the output is obtained as the maximum value of the probability of the input linguistic term, as given in Equation 11. Then, for similar output linguistic term obtained from different rules, the maximum value is considered.

$$\mu^{A,B} = \max(\mu^A(x), \mu^B(x)) \qquad (11)$$

**Table 1.** Examples of the set of fuzzy rules

| | Rules | | |
|---|---|---|---|
| IF | *q* is empty | THEN | *Dp* is zero |
| IF | *q* is low | THEN | *Dp* is zero |
| IF | *q* is moderate | AND THEN | *avg* is low *Dp* is low |
| IF | *q* is moderate | AND THEN | *avg* is moderate *Dp* is low |
| IF | *q* is moderate | AND THEN | *avg* is high *Dp* is medium |
| IF | *q* is full | AND THEN | *avg* is low *Dp* is medium |
| IF | *q* is full | AND THEN | *avg* is moderate *Dp* is high |
| IF | *q* is full | AND THEN | *avg* is high *Dp* is high |

### 3.4.4 Defuzzification

*Defuzzification,* the final step generates a crisp value for the output linguistic variable from its linguistic term. One of the popular defuzzification techniques is the center of gravity (COG) method. COG finds the point located on the center of all the output linguistic terms. Given that multiple output linguistic terms are obtained as multiple rules are anticipated, each with a probability value, the final output crisp result is obtained as calculated in Equation 12.

$$COG = \frac{\sum_{x=a}^{b} \mu^A(x)x}{\sum_{x=a}^{b} \mu^A(x)} \quad (12)$$

where the upper term represents the value of the probability of the output linguistic term x multiplied by all the values covered by this term, e.g.: 0.0, 0.1 and 0.2. The lower term represents the number of values multiplied by the probability.

**Table 2.** Complete set of fuzzy rules formed in a table

|     |          | q |   |   |   |
| --- | -------- | --- | --- | --- | --- |
|     |          | **empty** | **low** | **moderate** | **full** |
| **avg** | **low** | zero | zero | low/ moderate/ | low/ moderate/ high |
|     | **moderate** | zero/ low/ moderate | zero/ low/ moderate | low/ moderate/ high | low/ moderate/ high |
|     | **high** | zero/ low/ moderate/ high | zero/ low/ moderate/ high | moderate/ high | low/ moderate/ high |

### 3.4.5 Calculate Final Dropping Probability

As a final and post fuzzy inference process, the output of the fuzzy inference process is then merged with the delay measure to produce the final dropping probability. The final dropping probability is calculated as given in Equation 13.

$$D_p = D_p' + w_D (D_{Esti}) \quad (13)$$

where $D_p'$ is the output of the fuzzy inference process and $D_p$ is the final dropping probability.

## 4. Results

The simulation of the network process is implemented using one of the well-known approaches called discrete time queue [6]. The discrete time queue tracks, measures and evaluates the status of the network and its resources at equal time intervals called slots. At each slot, either a packet arrive event or departed even or both events may occur. Two subsequent packets arrival without departure makes two time slots and so on [2, 6]. Several methods have been modeled and tested using discrete-time queues [6, 22-25]. The other approach, called continuous model, measures and evaluates the network performance periodically with equally length periods. However, this approach does not efficiently address the events of packet arrival and departure accurately. As the AQM is based on calculating $D_p$ with each arrival packet (event based), discrete time queue was chosen to verify the proposed work.

The proposed and compared methods are simulated under the discrete time queue in Java. The programming code is developed in NetBeans 7.4 IDE. The results are obtained in a machine with Windows 7 operating system, in Core i7 2400GHz and 6 GB RAM. The parameters used in the experiments for the proposed and compared methods are given in Table 3 as recommended in RED and ERED.

Note that these values are used with all the methods for fair comparison. Other parameters that are used by the proposed methods are determined experimentally.

The experiment is conducted as follows: The parameters, as given in Table 3, is initialized first. Then, in each time slot, a packet maybe generated according to the probability of packet arrival and is sent to the queue. The packet as generated and sent, maybe lost if the queue is full, or it might be dropped or queued according to a decision made based on the implemented method and the generated $Dp$ value. In the same time slot, a packet maybe departed according to the probability of packet departure. These processes are repeated in each time slot.

Finally, the results are collected and reported. These processes are illustrated in Fig. 3.

**Table 3.** Parameter settings

| Parameter | Values |
|---|---|
| Probability of packet arrival | 0.33-0.93 |
| Probability of packet departure | 0.5 |
| Number of slots | 2,000,000 |
| Warm-up | 800,000 |
| Router buffer capacity | 20 |
| Queue weight | 0.002 |
| $max_p$ | 0.1 |
| $min_{th}$ | 3 |
| $max_{th}$ | 9 |

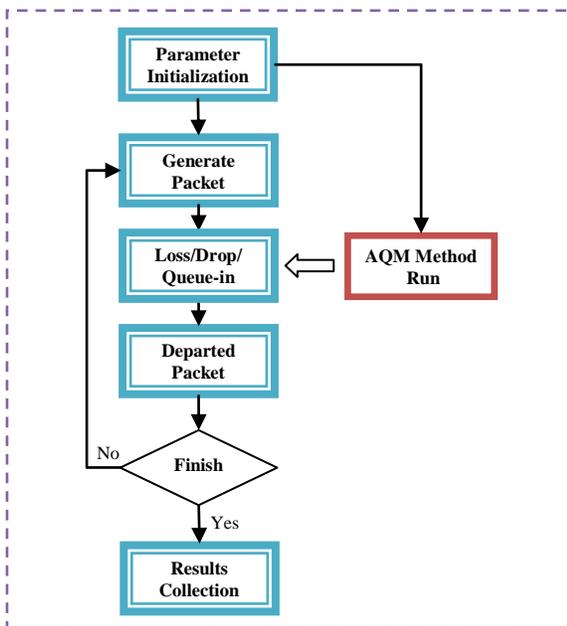

**Fig. 3.** Flowchart of the experiments

### 4.1 Parameter settings

Besides the common parameters introduced above, the proposed method involve other parameters that are: The delay weight, which is the significance given to the delay in the overall *Dp* calculation. The weight that is used in the calculation of the arrival rate, this weight is used to balance between the previous and current arrival values to produce weighted average value for the estimated arrival rate. The weight that is used in the calculation of the departure rate. These parameters are set up empirically side by side with the fuzzy sets and memberships values that are used in the Fuzzy Inference Process (FIP). The values for the weights that are tested and the final weights set up are given in Table 4. While the final sets and memberships are given in Table 5. The FIP values are set up in trial and error approach.

**Table 4.** Empirical weight settings

| Parameter | Tested Values | Final Value |
|---|---|---|
| Delay Weight | 1.0, 0.9, 0.8, .., 0.1, 0.09, 0.08, ..0.01 | 0.05 |
| Arrival Weight | 1.0, 0.9, 0.8, .., 0.1, 0.09, 0.08, ..0.01 | 0.2 |
| Departure Weight | 1.0, 0.9, 0.8, .., 0.1, 0.09, 0.08, ..0.01 | 0.2 |

### 4.2 Results

The overall results of the compared and proposed methods are given in Table 6. The results is compared under a heavy congestion, where the arrival probability is 0.93, while the departure probability is 0.5. As given in the table, the number of packet arrival and the number of packet departure for used with the execution of each method are listed first. Note that, the numbers is not totally equal because the arrival depends on a probability value which is 0.93 of the 1,200,000 time slots. However, the packet arrival is almost equal. As for the packet departure, not all packets arrived will be departed as these packets are subject to packet loss and packet dropping, thus differs from one method to another. The first 800,000 time slots out of the total times slots 2,000,000, which are used in the experiments are for warm-up. The measures and all factors in warm-up period are simply discarded.

Table 5. Empirical FIP settings

| Parameter | Tested Values | Final Value |
|---|---|---|
| Queue Set | 2, 3, 4, 5, 6 Linguistic Variables | 4 Linguistic Variable: {zero, low, moderate, full} |
| Queue Membership | {0,0,0.6,0.7},{0.6,0.7,0.7,0.8},{0.7,0.8,0.8,0.9},{0.8,0.9,0.9,1.0} | |
| Average Queue Length Set | 2, 3, 4, 5, 6 Linguistic Variables | 3 Linguistic Variable: { low, moderate, high} |
| Average Queue Length Membership | {0,0,0.6,0.7},{0.6,0.8,0.8,0.9},{0.8,0.9,0.9,1.0} | |
| Dropping Probability | 2, 3, 4, 5, 6 Linguistic Variables | 4 Linguistic Variable: {zero, low, moderate, high} |
| Dropping Probability Membership | {0,0,0.3,0.4},{0.3,0.4,0.5,0.6},{0.5,0.6,0.8,0.9},{0.8,0.9,1.0,1.0} | |
| Rules | IF $q$ is empty THEN $Dp$ is zero | |
| | IF $q$ is low THEN $Dp$ is zero | |
| | IF $q$ is moderate AND $avg$ is low THEN $Dp$ is low | |
| | IF $q$ is moderate AND $avg$ is moderate THEN $Dp$ is low | |
| | IF $q$ is moderate AND $avg$ is high THEN $Dp$ is moderate | |
| | IF $q$ is full AND $avg$ is low THEN $Dp$ is moderate | |
| | IF $q$ is full AND $avg$ is moderate THEN $Dp$ is high | |
| | IF $q$ is full AND $avg$ is high THEN $Dp$ is high | |

The number of packet loss and packet drops are then given and the total number of aggregating these values are given as packet missed. Note that, if a similar number of packets missed in all methods, it does not means that there performance are equal, because dropping a packet is better than losing it, as simple fact in network performance evaluation. Subsequently, a method that has ability to predict congestion in early stage and drops a packet is highly better than the one that will finally queued-up and starts losing packets. The aim of showing the overall packet missed is to show that having more packet dropping is an advantage in case that the packet will finally be dropped, while dropping packets unnecessarily when there is no risk to loss packets is disadvantage.

Table 6. Performance comparison with arrival rate 0.93

| | RED | ERED | Hybrid | Fuzzy |
|---|---|---|---|---|
| #Packet arrived | 1115557 | 1115802 | 1115802 | 1115592 |
| #Packet depart. | 598376 | 599997 | 599997 | 599275 |
| #Packet Loss | 323045 | 69893 | 3385 | 0 |
| #Packet dropped | 194136 | 445914 | 511787 | 516318 |
| #Packet missed | 517181 | 515807 | 515172 | **516318** |
| delay | 33.63675 | 32.27664 | 23.2277 | **21.46638** |

As noted, the proposed Hybrid-ERED improves packet loss measure by dropping more packets. This dropping cannot be considered as disadvantage, because finally the overall missed packets for ERED and Hybrid-ERED is equal. Subsequently, Hybrid-ERED drops packets in-order to prevent the loss, which will surely occur in this case as the arrival rate is higher than departure rate. Moreover, the delay is enhanced significantly by the proposed Hybrid-ERED.

As for the fuzzy based method, it is notably that using FIP not only eases the parameter settings, which was the purpose of using it, but also enhances the results in term of packet loss and delay. Subsequently, the proposed method in its parametric and FIP-based forms performs highly better than ERED.

The results of the compared and proposed methods where the arrival probability is 0.66,

0.5 and 0.33, while the departure probability is 0.5 are given in Table 7, Table 8 and Table 9. As noted the number of packet arrivals in the following tables are decreased as the arrival probability decreased.

As noted, when the arrival probability decreased from 0.93 to 066 (Table 7 and Table 8), the performance of the Hybrid is better than of the fuzzy method in term of the packet dropping. Packet loss and delay are still better in the fuzzy method. However, dropping packets unnecessary is disadvantage. Thus, the best performance can be awarded here to the hybrid method. If the hybrid method is compared to ERED and RED, fuzzy method losses less packets and has less overall packet missed. Overall, the proposed method is its parametric form performs better than ERED, when arrival probability is equal to 0.66.

**Table 7.** Performance comparison with arrival rate 0.66

|  | RED | ERED | Hybrid | Fuzzy |
|---|---|---|---|---|
| #Packet arrived | 791269 | 789264 | 788967 | 791685 |
| #Packet depart. | 595964 | 601946 | 602121 | 566053 |
| #Packet Loss | 77469 | 8765 | 3385 | **0** |
| #Packet dropped | 117828 | **178555** | 183463 | 225632 |
| #Packet missed | 195297 | 187320 | **186848** | 225632 |
| delay | 28.37491 | 24.1288 | 20.237 | **9.45644** |

When the arrival probability decreased to be 0.5 which is equal to the departure probability, this case means no congestion occurs and the congestion control method should not drop packets excepts in some cases. In Table 8 it is noted that the performance of the hybrid method is better than of the fuzzy method in term of packet dropping. Packet loss and delay are still better in the fuzzy method. If compared to ERED and RED, the hybrid method losses less packets and has less overall packet missed. Overall, the proposed method in its parametric form performs better than ERED, when arrival probability is equal to 0.5.

When the arrival probability decreased to be 0.33 which is less than the departure probability, this case means no congestion occurs and the congestion control method should not drop packets. In Table 9, it is noted that the performance of all the compared methods are the same.

**Table 8.** Performance comparison with arrival rate 0.5

|  | RED | ERED | Hybrid | Fuzzy |
|---|---|---|---|---|
| #Packet arrived | 601386 | 602113 | 601959 | 597847 |
| #Packet depart. | 575794 | 581530 | 581364 | 460056 |
| #Packet Loss | 5070 | 2055 | 2055 | **0** |
| #Packet dropped | 20534 | 18526 | 18526 | 29787 |
| #Packet missed | 25604 | **20581** | **20581** | 29787 |
| delay | 16.53937 | 17.85275 | 17.69034 | **8.66399** |

**Table 9.** Performance comparison with arrival rate 0.33

|  | RED | ERED | Hybrid | Fuzzy |
|---|---|---|---|---|
| #Packet arrived | 394363 | 394363 | 394363 | 395451 |
| #Packet depart. | 394363 | 394363 | 394363 | 394363 |
| #Packet Loss | **0** | **0** | **0** | **0** |
| #Packet dropped | **0** | **0** | **0** | **0** |
| #Packet missed | **0** | **0** | **0** | **0** |
| delay | 3.92946 | 3.92946 | 3.92946 | 3.92946 |

To give more depth on the performance of the proposed and compared methods, these methods are compared with different arrival probabilities, with mql, dropping, PL, missed, delay and throughput as given in Fig. 4.

As noted, fuzzy method trends to drop packets unnecessarily, unlike the parametric-based hybrid method. Subsequently, the parametric-based can be used when all the performance measures are demanded while the performance of the fuzzy is still superior in case that delay is a critical factor, because it produce superior delay minimization compares to the other methods. Finally, the throughput which depends on the number of packet departed and the utilization of queue, is almost equal for all methods excepts it is less for FIP-based form as it drops more packets.

## 5. Conclusion

In this paper, a method for congestion control was developed and tested. The proposed method extends a well known method called ERED in two ways: First, the proposed method extends ERED by considering the delay issues in the router buffer. The delay has been estimated using an equation with the inputs of estimated arrival and estimated departure rates. Subsequently, the extended method has transferred the original ERED method from being queue-based into queue-based and load-based (hybrid) method. Second, the proposed hybrid method has been implemented with a Fuzzy Inference Process (FIP) to ease the problem of parameter initialization.

Experimental validation has been implemented. The result shows that the proposed method in its two forms, parametric-based and fuzzy-based, has improved the performance of congestion control in terms of packet dropping, packet loss and delay. The performance was enhanced by decreasing both the packet loss and the queuing delay. The proposed work was proved its ability to overcome the limitations of the early method which are reducing delay and ease the parameter initialization problem.

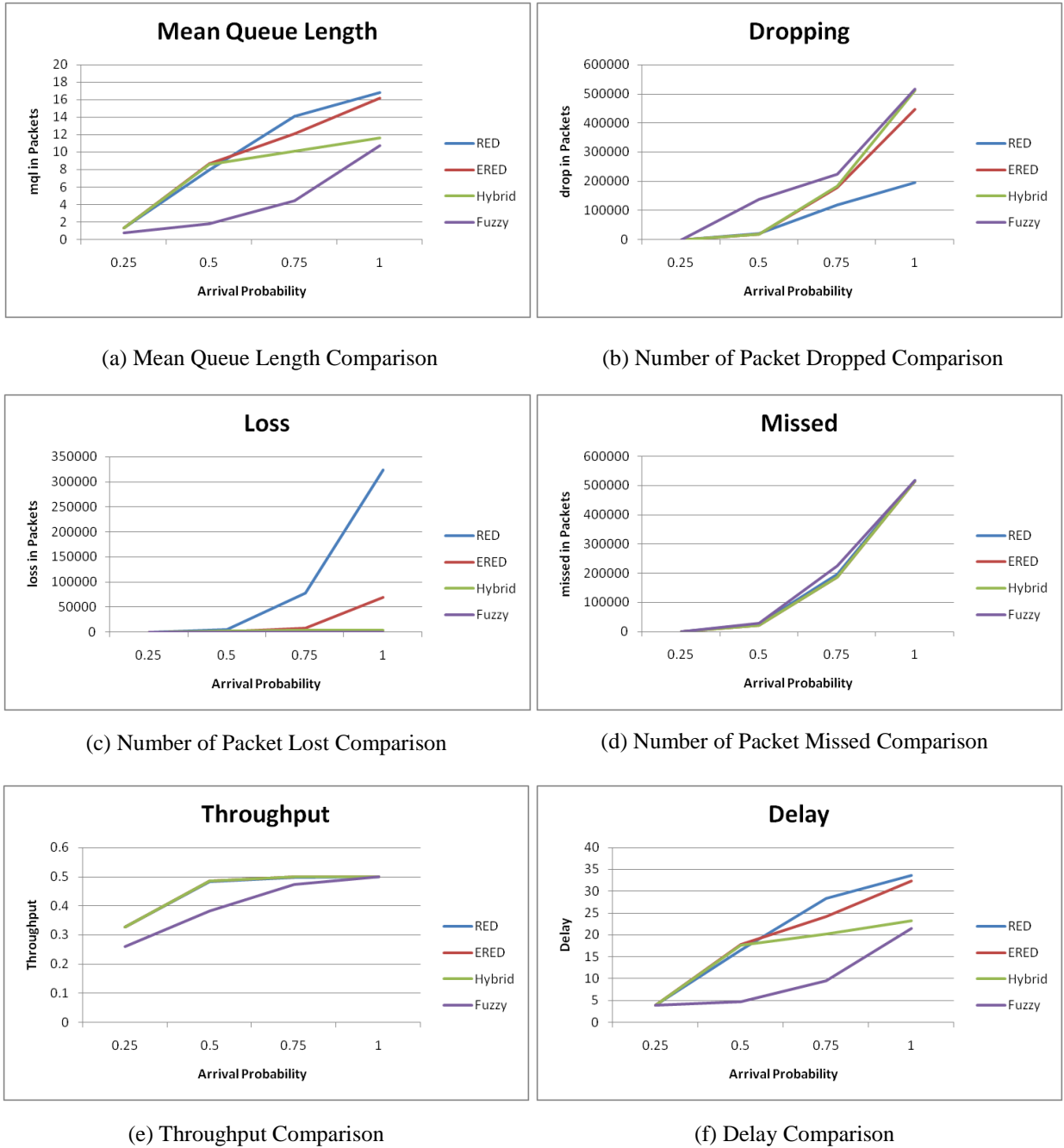

**Fig. 4.** Performance comparison

**References**

1. Lim, L.B., et al., *Controlling mean queuing delay under multi-class bursty and correlated traffic.* Journal of Computer and System Sciences, 2011. 77(5): p. 898-916.

2. Woodward, M.E., *Communication and Computer Networks: Modelling with discrete-time queues.* 1993: Wiley-IEEE Computer Society Press.

3. Chitra, K. and D.G. Padamavathi, *Adaptive CHOKe: An algorithm to increase the fairness in Internet Routers* Int. J. Advanced

Networking and Applications, 2010. 01(06): p. 382-386
4. Seifaddini, O., A. Abdullah, and a.H. Vosough, *RED, GRED, AGRED CONGESTION CONTROL ALGORITHMS IN HETEROGENEOUS TRAFFIC TYPES*, in *International Conference on Computing and Informatics*. 2013.
5. Abbasov, B. and S. Korukoglu, *Effective RED: An algorithm to improve RED's performance by reducing packet loss rate.* Journal of Network and Computer Applications, 2009. 32(3): p. 703-709.
6. Alfa, A.S., *Queueing Theory for Telecommunications*. 2010: Springer US.
7. Floyd, S. and V. Jacobson, *Random early detection gateways for congestion avoidance.* IEEE/ACM Trans. Netw., 1993. 1(4): p. 397-413.
8. Kwon, M. and S. Fahmy. *Comparison of load-based and queue-based active queue management algorithms*. in *ITCom 2002: The Convergence of Information Technologies and Communications*. 2002: International Society for Optics and Photonics.
9. Kunniyur, S. and R. Srikant, *End-to-end congestion control schemes: Utility functions, random losses and ECN marks.* Networking, IEEE/ACM Transactions on, 2003. 11(5): p. 689-702.
10. Lapsley, D. and S. Low. *Random early marking: an optimisation approach to Internet congestion control*. in *Networks, 1999. (ICON '99) Proceedings. IEEE International Conference on*. 1999.
11. Hollot, C.V., et al. *On designing improved controllers for AQM routers supporting TCP flows*. in *INFOCOM 2001. Twentieth Annual Joint Conference of the IEEE Computer and Communications Societies. Proceedings. IEEE*. 2001: IEEE.
12. Athuraliya, S., et al., *REM: active queue management.* Netwrk. Mag. of Global Internetwkg., 2001. 15(3): p. 48-53.
13. Chrysostomou, C., et al., *Fuzzy Explicit Marking for Congestion Control in Differentiated Services Networks*, in *Proceedings of the Eighth IEEE International Symposium on Computers and Communications*. 2003, IEEE Computer Society.
14. Negnevitsky, M., *Artificial Intelligence: A Guide to Intelligent Systems*. Second Edition ed. 2005, England.
15. Hiok, H. and B. Qiu. *Fuzzy logic target utilization and prediction for traffic control*. in *Global Telecommunications Conference, 2000. GLOBECOM '00. IEEE*. 2000.
16. Chrysostomou, C., et al., *Congestion control in differentiated services networks using Fuzzy-RED.* Control Engineering Practice, 2003. 11(10): p. 1153-1170.
17. Abdel-Jaber, H., et al. *Fuzzy logic controller of Random Early Detection based on average queue length and packet loss rate*. in *Performance Evaluation of Computer and Telecommunication Systems, 2008. SPECTS 2008. International Symposium on*. 2008.
18. Yaghmaee, M.H. and H. AminToosi, *A Fuzzy Based Active Queue Management Algorithm.* Computer Department, Ferdowsi University of Mashhad, Faculty of Engineering, Mashad, 2003: p. 458-462.
19. Zadeh, L.A., *Fuzzy sets.* Information and Control, 1965. 8(3): p. 338-353.
20. Klir, G.J., *Fuzzy logic.* Potentials, IEEE, 1995. 14(4): p. 10-15.
21. Allen, A.O., *Probability, statistics, and queueing theory*. 1990: Academic Press.
22. Abdel-Jaber, H., et al., *Performance evaluation for DRED discrete-time queueing network analytical model.* J. Netw. Comput. Appl., 2008. 31(4): p. 750-770.
23. Henderson, W., et al., *Closed queueing networks with batch services.* Queueing Systems, 1990. 6(1): p. 59-70.
24. Baklizi, M., et al., *Dynamic stochastic early discovery: a new congestion control technique to improve networks performance.* ICIC International, 2013. 9(3): p. 1113-1126.
25. Baklizi, M., et al., Fuzzy logic controller of gentle random early detection based on average queue length and delay rate. International Journal of Fuzzy Systems, 2014. 16(1): p. 9-19.